\documentclass{PoS}
\def\inlinetilde{\lower0.8ex\hbox{$\,\widetilde{}\,$}}
\def\chpt{\raise0.4ex\hbox{$\chi$}PT}
\def\schpt{S\raise0.4ex\hbox{$\chi$}PT}

\def\midtild{\lower1ex\hbox{$\tilde{}$}}

\def\leftvec{{\raise1.5ex\hbox{$\leftarrow$}\kern-.85em}}
\def\half{{\scriptstyle \raise.2ex\hbox{${1\over2}$}}}
\def\threehalves{{\scriptstyle \raise.15ex\hbox{${3\over2}$}}}
\def\third{{\scriptstyle \raise.15ex\hbox{${1\over3}$}}}
\def\third{{\scriptstyle \raise.15ex\hbox{${1\over3}$}}}
\def\twothirds{{\scriptstyle \raise.15ex\hbox{${2\over3}$}}}
\def\fourth{{\scriptstyle \raise.15ex\hbox{${1\over4}$}}}
\def\gtwid{{\,\raise.3ex\hbox{$>$\kern-.75em\lower1ex\hbox{$\sim$}}\,}}
\def\ltwid{{\,\raise.3ex\hbox{$<$\kern-.75em\lower1ex\hbox{$\sim$}}\,}}

\def\circle{{\Large{\raise-0.15ex\hbox{$\circ$}}}}
\def\sqar{{\raise-0.1ex\hbox{$\Box$}}}   %scales with fonts
\def\et{{\it et al.}}

\def\nsection#1 #2{\leftline{\rlap{#1}\indent\relax #2}}

\def\prd#1{Phys.\ Rev.\ D {\bf #1}}

\usepackage[latin1]{inputenc}
\usepackage{epsf}
\usepackage{graphicx,psfrag,color,pstcol,pst-grad}
\usepackage{amsmath,amssymb}
\usepackage{latexsym}
\usepackage{calc}

\def\as{\alpha_s}
\def\inlinetilde{\lower0.8ex\hbox{$\,\widetilde{}\,$}}
\def\chpt{{\color{salmon}\raise0.4ex\hbox{$\chi$}PT}}
\def\schpt{{\color{salmon}S\raise0.4ex\hbox{$\chi$}PT}}

\def\bluecircle{{\Large{\color{blue}\raise-0.15ex\hbox{$\circ$}}}}
\def\redcircle{{\Large{\color{red}\raise-0.15ex\hbox{$\circ$}}}}
\def\bluebox{{\color{blue}\raise-0.1ex\hbox{$\Box$}}}
\def\blackbox{{\color{black}\raise-0.1ex\hbox{$\Box$}}}

\PoS{PoS(LAT2005)203}

\title{Onium Masses with Three Flavors of Dynamical Quarks }

\ShortTitle{Onium Masses with Three Flavors of Dynamical Quarks}

\author{\speaker{Steven Gottlieb},  %\thanks{A footnote may follow.}, 
	L.~Levkova\\
        Department of Physics, Indiana University, Bloomington, Indiana 47405, USA\\
        E-mail: \email{sg@indiana.edu, llevkova@indiana.edu}}

\author{M.~Di Pierro\\
	School of Computer Science, Telecommunications and Information Systems,
        DePaul University, Chicago, Illinois 60604, USA\\
        E-mail: \email{MDiPierro@cs.depaul.edu}}
\author{A.~El-Khadra, D.~Menscher\\
	Physics Department,
        University of Illinois, Urbana, Illinois 61801, USA\\
        E-mail: \email{axk@uiuc.edu, menscher@uiuc.edu}}
\author{
A.S.~Kronfeld, P.B.~Mackenzie, J.~Simone\\
        Fermi National Accelerator Laboratory, Batavia, Illinois 60510, USA\\
        E-mail: \email{ask@fnal.gov, mackenzie@fnal.gov, simone@fnal.gov}}

\abstract{
We have greatly extended an earlier calculation of the charmonium spectrum
on three flavor dynamical quark ensembles by using more recent ensembles 
generated by the MILC collaboration.  The heavy quarks are
treated using the Fermilab formulation.  The charmonium state masses are
in reasonable agreement with the observed spectrum; however, some of the
spin splittings may still be too small.}

\FullConference{XXIIIrd International Symposium on Lattice Field Theory\\
		 25-30 July 2005\\
		 Trinity College, Dublin, Ireland}

\begin{document}

\section{Introduction}
Calculating the spectrum of onium states is a significant challenge
for lattice gauge theory.  
A number of levels can be studied for both charm and bottom quarks.  
However, dealing with heavy quarks requires special care
\cite{NRQCD92,kkm}.
Using improved staggered sea quarks \cite{ASQTAD}, it is possible to 
reproduce many 
of the most important features of the spectrum \cite{PRL}, which had not
been done in the quenched approximation.
This paper updates our work presented at Lattice 2003 \cite{onium2003}.

\section{Calculational Details}

Ensembles for three lattice spacings were provided by the MILC Collaboration
\cite{MILCLATS}:
$a\! \approx\! 0.18\;$fm (``extra-coarse''),
$a\! \approx\! 0.12\;$fm (``coarse''),
and $a \! \approx\! 0.086\;$fm (``fine'').  
(See Table \ref{latticetable}.)
For the extra coarse $am_q$ / $am_s$ = 0.6, 0.4, 0.2 and 0.1; 
for the coarse lattice, we also have 0.14, but we have only 
analyzed two values 0.4 and 0.2 for the fine lattice.
From 400 to 600 configurations have been analyzed in most ensembles.
The most notable exception is the coarse ensemble with $am_q$ / $am_s =0.1$.
For each of the lattice spacings, the scale of each ensemble
with different sea quark masses was kept approximately fixed
using the length $r_1$ \cite{SOMMER,MILC-POTENTIAL} from 
the static quark potential.
%%We reduce the statistical fluctuations in $r_1/a$ by fitting to a
%%smooth function and using fit values.  
The absolute scale from the $\Upsilon$ $2S$--$1S$ splitting was
determined by the HPQCD/UKQCD group \cite{DAVIES,PRL} on most of 
our ensembles implying  $r_1 = 0.318(7)\;$fm.  

%%\section{Ensembles analyzed}

\begin{table}[bth]
\begin{center}
\begin{tabular}{|c|c|c|c|c|c|}
\hline
%%%\noalign{\vspace{0.15cm}}
$am_q$ / $am_s$  & \hspace{-1.0mm}$10/g^2$ & size & volume & config. & $a$ (fm) \\
\hline
0.0492  / 0.082   & 6.503 & $16^3\times48$ & $(2.8\;{\rm fm})^3$ & 401 & 0.178\\
0.0328  / 0.082   & 6.485 & $16^3\times48$ & $(2.8\;{\rm fm})^3$ & 331 & 0.177\\
0.0164  / 0.082   & 6.467 & $16^3\times48$ & $(2.8\;{\rm fm})^3$ & 645 & 0.176\\
0.0082  / 0.082   & 6.458 & $16^3\times48$ & $(2.8\;{\rm fm})^3$ & 400 & 0.176\\
\hline
0.03  / 0.05   & 6.81 & $20^3\times64$ & $(2.4\;{\rm fm})^3$ & 559 & 0.120\\
0.02  / 0.05   & 6.79 & $20^3\times64$ & $(2.4\;{\rm fm})^3$ & 460 & 0.120\\
0.01  / 0.05   & 6.76 & $20^3\times64$ & $(2.4\;{\rm fm})^3$ & 593 & 0.121\\
0.007  / 0.05   & 6.76 & $20^3\times64$ & $(2.4\;{\rm fm})^3$ & 403 & 0.121\\
0.005  / 0.05   & 6.76 & $24^3\times64$ & $(2.9\;{\rm fm})^3$ &  136 & 0.120 \\
\hline
0.0124  / 0.031   & 7.11 & $28^3\times96$ & $(2.4\;{\rm fm})^3$ & 261 & 0.0863 \\
0.0062  / 0.031   & 7.09 & $28^3\times96$ & $(2.4\;{\rm fm})^3$ & 472  & 0.0861\\
\hline
\end{tabular}
\end{center}
\caption{Ensembles used in this calculation.}
\label{latticetable}

\end{table}
%%%
%%%Table 1.  Ensembles of configurations.
%%%Lattice spacings
%%%come from the smoothed $r_1/a$ values and $r_1=0.318\;$fm.
%%%The top group of four ensembles is called ``extra coarse.''  The next group 
%%%of five ensembles is called ``coarse'' and the last pair denoted ``fine.''

%%\section{Lattice action}

We use the Asqtad improved staggered sea quark action that
has errors of $O(\as a^2)$.
The improved gluon action has errors of $O(\as^2 a^2)$.
For the heavy valence quarks, we  use the
Sheikholeslami-Wohlert action \cite{sw} (which has $O(\as a)$ errors)
with the Fermilab interpretation \cite{kkm}.
To compute heavy quark propagators, we use point and smeared sources
and sinks.  The smearing approximates 1S or 2S wavefunctions.
At the sink, spatial momentum $2\pi/(La) [ p_x, p_y, p_z ] $ is given to
the onium state.  We restrict the range of $p$ such that $\sum p_i^2 \le 9$.

%%\section{Fitting the propagators}

%%To fit the propagators we make use of a Baysian fitter.  
To find the onium masses,
we fit two channels simultaneously for the zero momentum states.  
A delta function and a 1S smearing wave function are used as 
the source and sink.
The ground state and up to three excited states are included in the fit.
The minimum and maximum distance from the source are varied, and the best
fit is selected based on the confidence level and size of error in the ground
state and first excited state masses.
%%We have not attempted to hand pick fits (with a few exceptions), 
%%nor have we added a systematic error based upon varying the choosen fit.
After choosing the fit range, 250 bootstrap samples 
are generated to provide an error estimate.

%%\section{Tuning the quark mass}

\begin{figure}[th]
\begin{center}
\includegraphics[width=7cm]{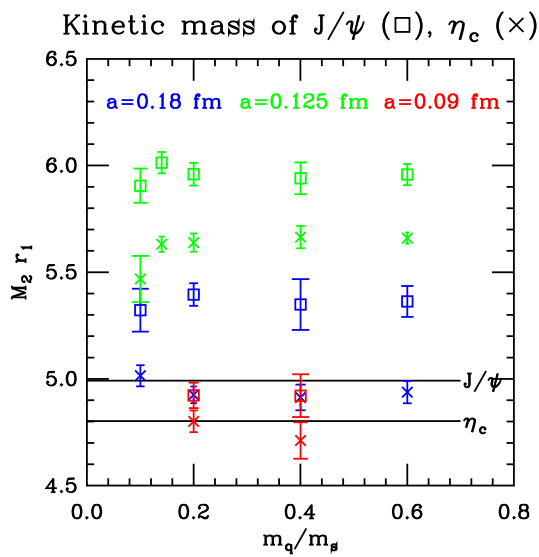}
\end{center}
\caption{The kinetic masses of $J/\psi$ and $\eta_c$ on each ensemble
plotted as a function $m_q/m_s$ the light sea to strange quark mass ratio.
Masses are in units of $r_1$.  The physical masses are shown as lines.}
\label{kineticmassfig}
\end{figure}

We must tune the hopping parameter $\kappa$ to the charm or bottom mass.
For each lattice spacing, we select a sea quark mass independent value for
$\kappa$.  The tuning is done on an ensemble with small sea quark mass.
In fact, as this project was done in conjuction with a study of heavy-light
mesons, the tuning was done for the $D_s$ mass.  The precision of that tuning
was only about 8\%.
Because of lattice artifacts that arise for heavy states, we distinguish
between the rest mass $aM_1$ and the kinetic mass $aM_2$.  
We use
$\kappa=0.120$, 0.119 and 0.127 on the extra coarse, coarse and fine ensembles, 
respectively.
%%We saw above that $\kappa=0.120$ was too heavy for the extra coarse
%%ensembles.  It follows that $\kappa=0.119$ on the coarse ensembles, 
%%which are closer to the continuum limit will be even heavier.
The imprecision of our tuning is immediately seen 
in Fig.~\ref{kineticmassfig}.

%%\section{Sea quark mass dependence}

%%\section{Rest masses, states}

The kinetic masses have two disadvantages:
%%they require careful tuning of the charm or bottom quark mass, and
their statistical errors are large compared to those of the rest masses,
and the pattern of systematic errors is more subtle \cite{ASK96}.
However, for level splittings, a large discretization effect in the quark's rest
mass drops out of the energy differences of hadron rest masses \cite{ASK00}.
So, having tuned to approximately the right charm mass, we will now
consider splittings based upon the rest masses of the various states.
These states have been studied:
	    $\eta_c (1S)$, $\eta_c (2S)$,
	    $\psi (1S)$, $\psi (2S)$,
	    $h_c (1P)$,
	    $\chi_{c0} (1P)$ and
	    $\chi_{c1} (1P)$.
The $\chi_{c2} (1P)$ is also under study 
%using a new program that implements
with
a nonrelativistic $P$-wave source \cite{DAVIES94}.
Currently, results for $\chi_{c2} (1P)$ are only available 
on one extra coarse ensemble. %%% with $\beta=6.503$ and $m_q=0.0492$.
We also use the spin-averaged mass, {\it e.g.}, 
$\overline {1S} = [ 3 M_{\psi(1S)} + M_{{\eta_c} (1S)} ]/4 $ 
to display some of the splittings in the spectrum.

%%\section{Chiral extrapolation}
\section{Results}

For each lattice spacing, we plot the splittings as a function of the mass of
the light sea quarks.
A linear chiral fit is made and the splitting is extrapolated to the physical
value of $\hat m = (m_u + m_d)/2$ where the lattice spacing dependent value
of $\hat m$ is determined from analysis of $\pi$ and $K$ meson decays
constants \cite{STRANGE-MASS}.
The light meson decay constant analysis has not yet been completed on the 
extra coarse ensembles, 
so the value of $\hat m$ used there is only a rough estimate.

%%\section{Spin-averaged 2S--1S splitting}

\begin{figure}[th]
\setlength{\tabcolsep}{5mm}
\begin{tabular}{cc}
\includegraphics[width=6.5cm]{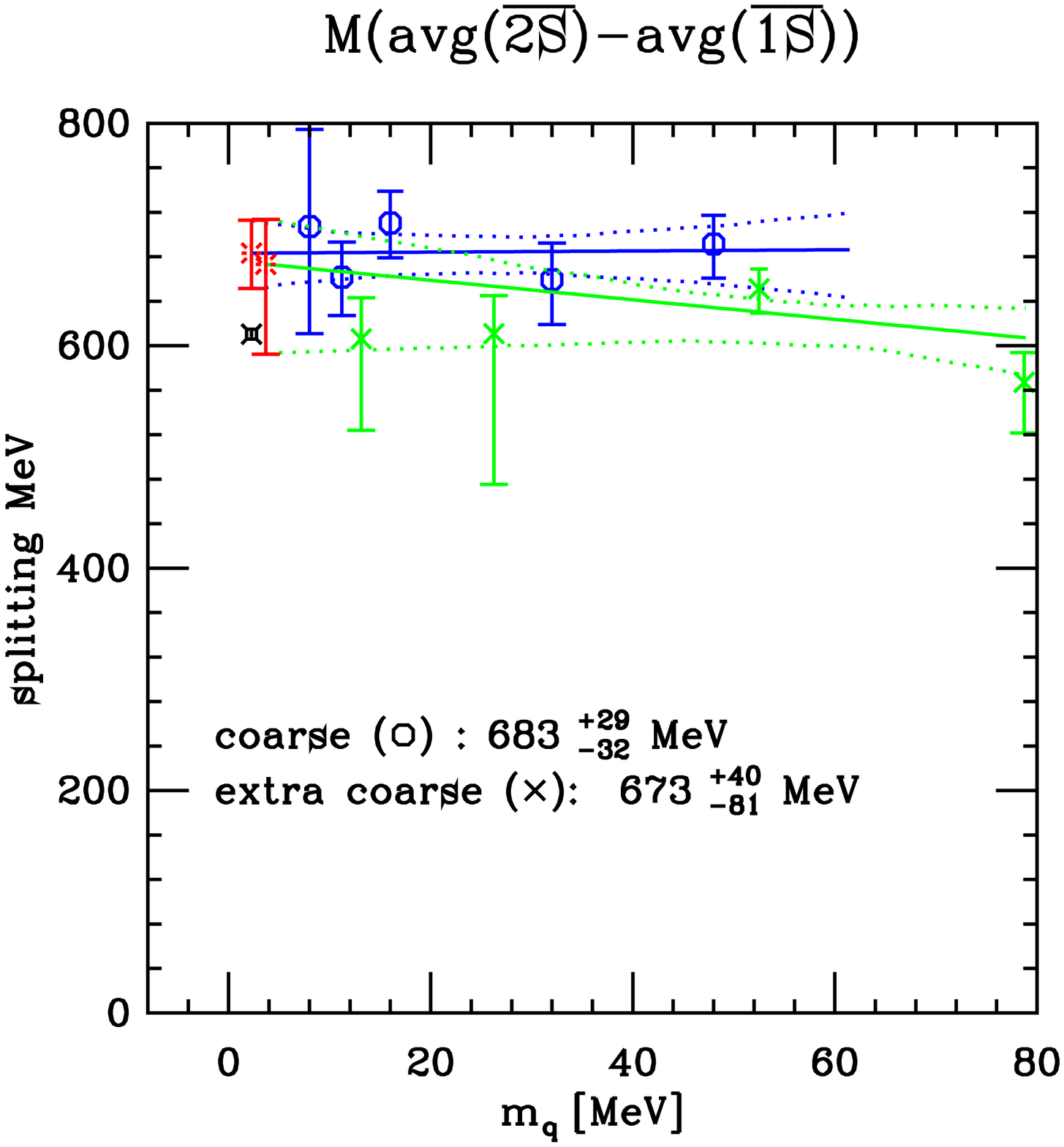}
&
\includegraphics[width=6.5cm]{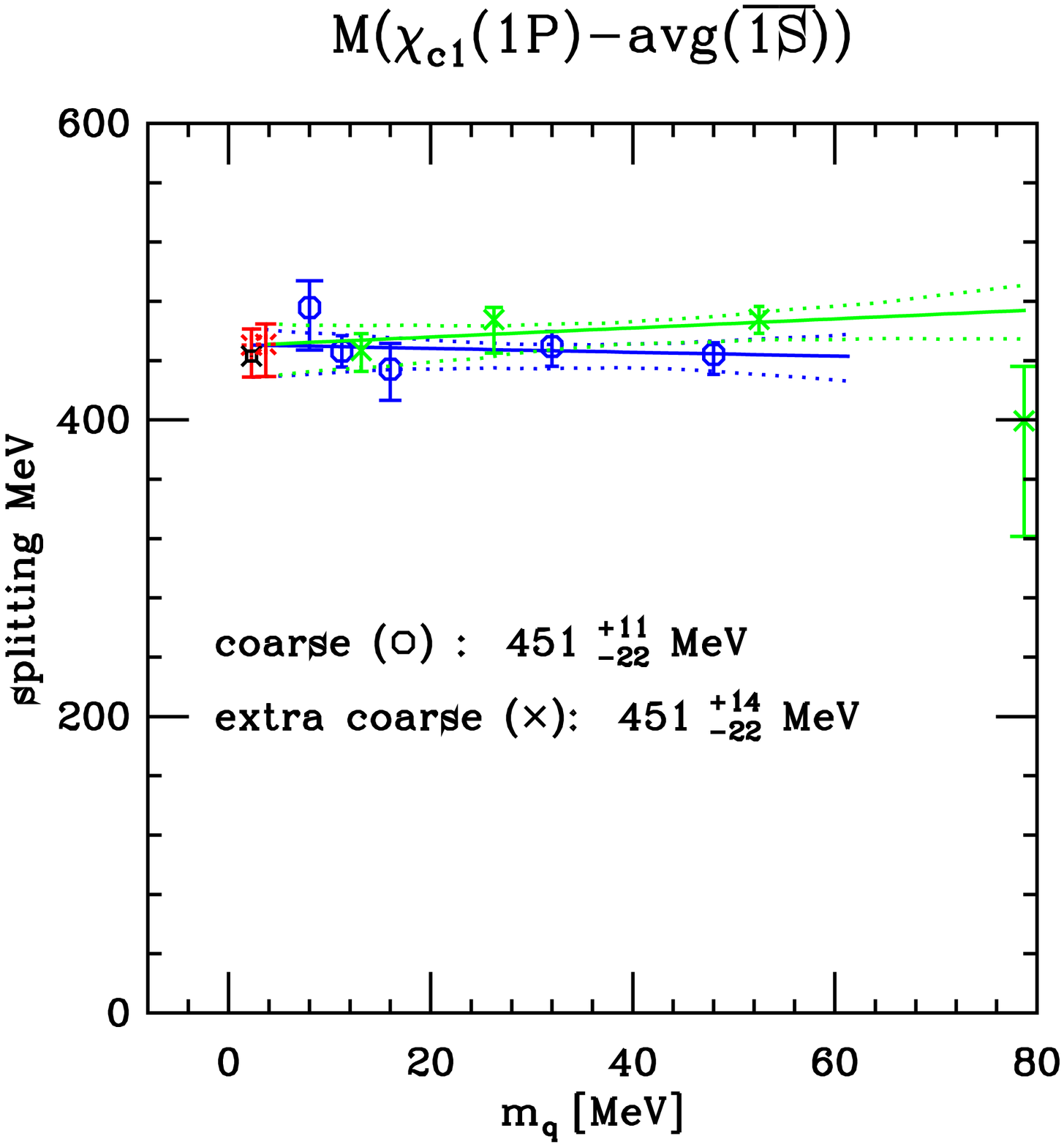}\\
\end{tabular}

\caption{(left) The chiral extrapolation of the spin-averaged splitting 
between the 2S and 1S states on the extra coarse and coarse ensembles.
The extrapolated values are shown in red, and the physical value in black.
}
\label{spinavgfig}
\caption{(right) The splitting 
between the $\chi_{c1}$ (${}^3P_1$) and spin-averaged 1S states 
on the extra coarse and coarse ensembles.
}
\label{chiorbitalfig}
\end{figure}

\begin{figure}[th]
\begin{tabular}{cc}
\includegraphics[width=6.5cm]{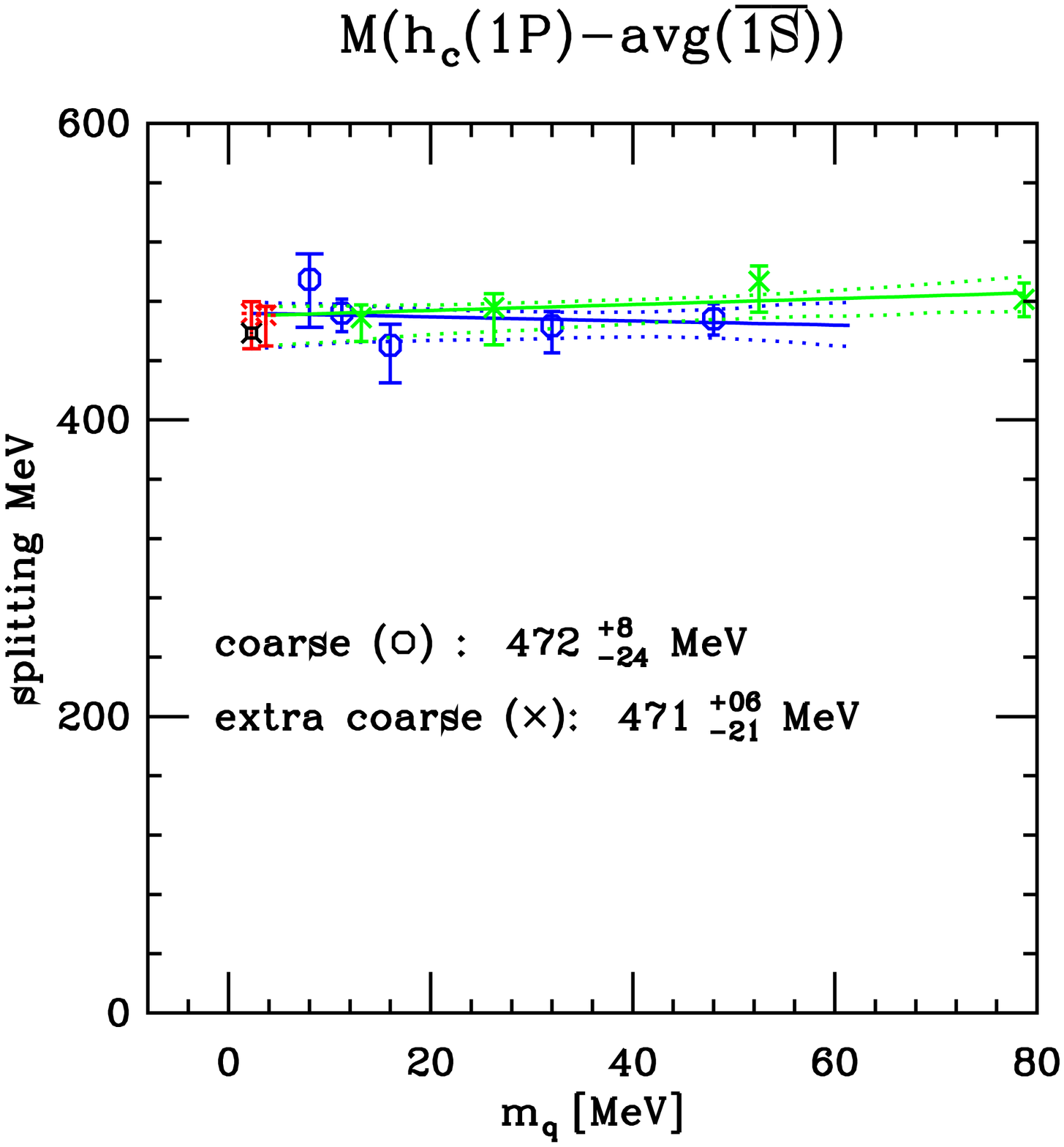}
&
\includegraphics[width=6.5cm]{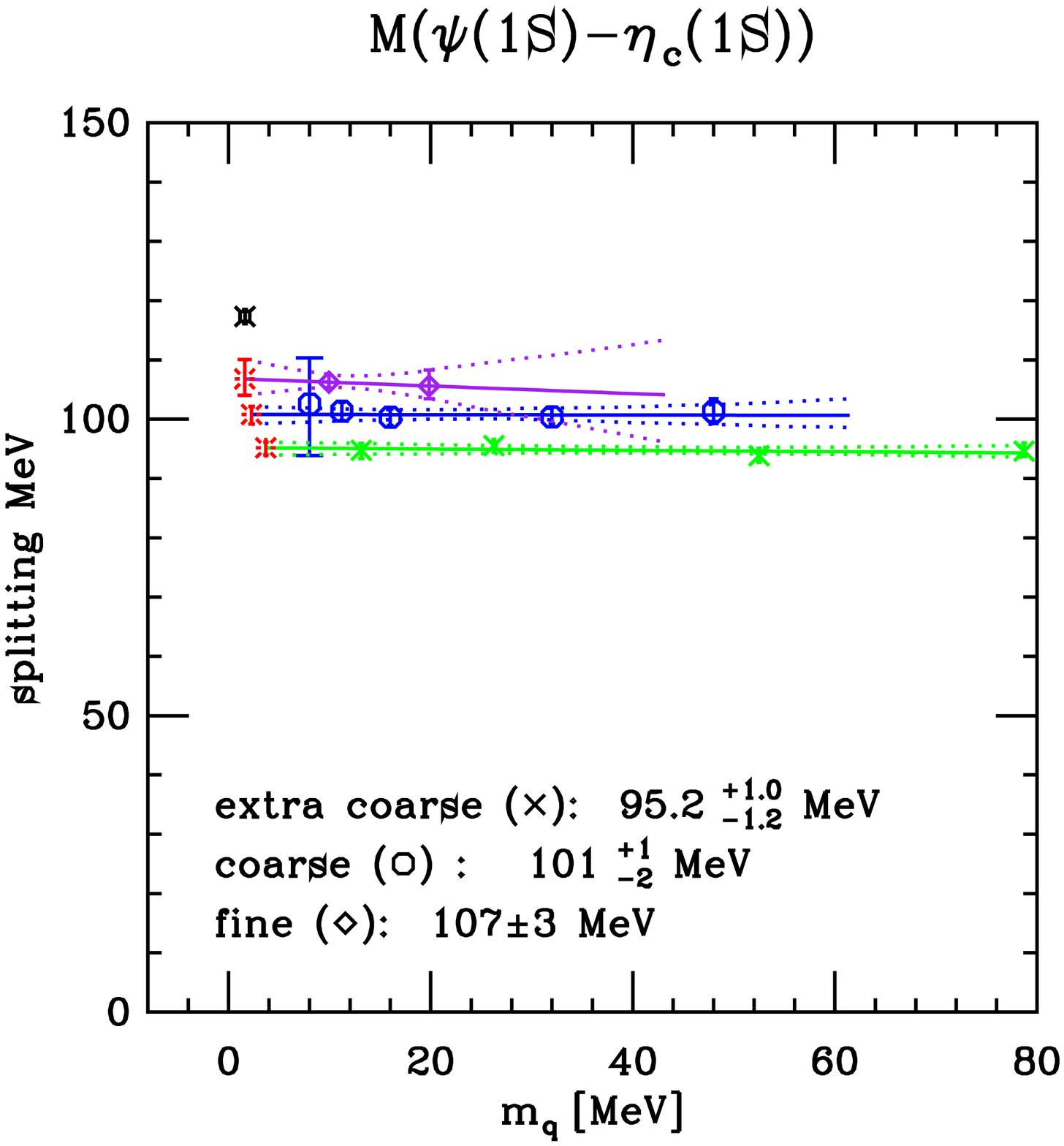}\\
\end{tabular}
\caption{(left) Same as Fig.~3,   %ABSOLUTE FIGURE REFERENCE
except for the $h_c$ (${}^1P_1$).}
\label{hcorbitalfig}
\caption{(right) Splitting between $J/\psi$ and $\eta_c$ for all three
lattice spacings.  There are only two ensembles for the fine lattice spacing,
shown in purple.}
\label{psietacfig}
\end{figure}

Within our current statistical uncertainties, we see reasonable agreement
with the experimental value of the splittings of the
spin-averaged 2S and 1S levels.  The coarse value is about 2
$\sigma$ high.  (See Fig.~\ref{spinavgfig}.)
As we do not yet have a full set of 
results for the $\chi_{c2}$, we cannot construct
the spin average of the 1P states.  Instead, we use the $\chi_{c1}$ and the
$h_c$.  
In nonrelativistic potential models, these two states are degenerate with each
other and the spin average.
The experimental splittings are well reproduced for these states.
(See Figs.~\ref{chiorbitalfig} and \ref{hcorbitalfig}.)

%%\section{Orbital splittings}

%%\includegraphics[width=7cm]{../sg_new_figs/chiral-chi_c11P-cc1S_coarse_xcoarse.ps}

%%\section{Spin splittings}
As seen in Fig.~\ref{psietacfig},
the spin splittings are too small.  For $J/\psi$ and $\eta_c$ it
amounts to about 10--22 MeV.  The splitting is 19\% too small for the 
extra coarse ensemble, 14\% too small for the coarse, 
and 9\% too small for the fine.
The splitting seems to systematically improve as the lattice spacing decreases.
We have not yet attempted a continuum extrapolation.

%%\item [{\color{magenta}$\bullet$}\ ]{} 
%%It had been hoped that adjusting terms in the heavy quark action
%%proportional to $\Sigma\cdot\vec B$ via perturbative matching might
%%resolve this issue \cite{nobes_lat03}.  However, it now appears that
%%these adjustments are too small to do so \cite{nobes_private}.
%%
%%\item [{\color{magenta}$\bullet$}\ ]{} 
%%An improved heavy quark action \cite{oktay} that includes terms such as
%%$\Sigma\cdot\vec B p^2 $ may be necessary to increase the spin splittings.

%%\section{Spectrum summary}

\begin{figure}[th]
\begin{center}
\includegraphics[width=8cm]{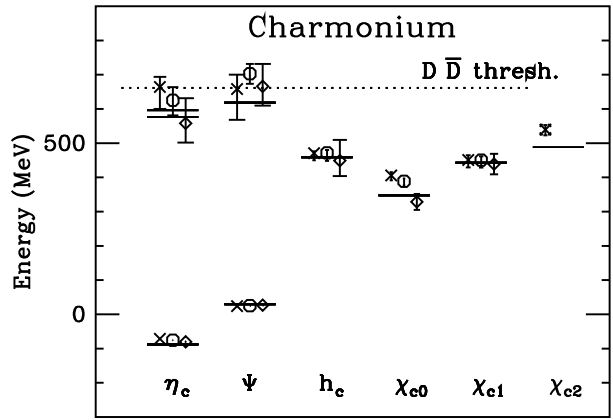}
\end{center}

\caption{Charmonium spectrum for all three lattice spacings
compared with experimental values.  Energy is offset so that zero represents
the spin-averaged 1S energy.  From left to right for each state, crosses,
octagons and diamonds are from the extra coarse, coarse and fine ensembles,
respectively.  The extra coarse $\chi_{c2}$ value without chiral extrapolation
is the fancy cross.}
\end{figure}

The overall agreement between this calculation with dynamical quarks and the
observed spectrum is very good.
The most obvious issue is the smallness of spin splittings, as seen in
the $J/\psi$ -- $\eta_c$ splitting, and the mass of the $\chi_{c0}$ state.
There is some evidence of improvement as the lattice spacing is reduced.

\section{Outlook}

There are several ways to improve this calculation in the near future:
		We can include another fine ensemble with $m_q=0.1 m_s$.  
		This more
		chiral ensemble is still being generated, but is far enough
		along that it would be worth starting the analysis.
		We also need to examine alternative fits to the ones that			were selected by our automated procedure.
		Fermilab/MILC are almost finished generating a new set of ensembles
		at a lattice spacing between extra coarse and coarse.
Production running on additional ensembles for the new $P$-wave
		code will be done.
We also plan to use heavier quarks to study 
bottomonium, which has already been studied on these
configurations using NRQCD \cite{DAVIES}.

In the longer term, MILC is generating new ensembles with $a\approx 0.06$ fm 
that should help us better understand the continuum limit.  
However, in the current
calculation, lattice spacing dependence does not seem very large compared with
statistical errors for most states.

We gratefully acknowledge the support of the Department of Energy.  
In addition, the Fermilab Computing Division, Indiana University (AVIDD cluster)
and National Center for Supercomputing Applications provided computing support.

%%%%%%%%%%%%%%%%%%%%%%%%%%%%%%%%%%%%%%%%%%%%%%%%%%%%%%%%%%%%%%

\end{document}